\documentclass[aps,prl,twocolumn,superscriptaddress]{revtex4-1}

\usepackage{mathrsfs,theorem,amssymb,wrapfig,amsmath}
\usepackage[]{graphicx}
\usepackage[normalem]{ulem}

\theoremstyle{definition}

\newcommand{\hil}{\mathcal{H}}
\newcommand{\hilb}[1][]{\mathcal{H}_{#1}}
\newcommand{\hilk}[1][]{\mathcal{K}_{#1}}

\newcommand{\ket}[1][]{| #1 \rangle}
\newcommand{\bra}[1][]{\langle #1 |}
\newcommand{\braket}[2]{\langle #1 | #2 \rangle}

\newcommand{\SU}[1][]{\mathrm{SU}(#1)}

\newcommand{\trace}[1][]{\mathrm{Tr}[{#1}]}

\newcommand{\concurrence}{\mathrm{C}}

\newcommand{\vac}{\mathrm{vac}}

\newcommand{\full}{\mathrm{full}}

\newcommand{\tr}{\mathrm{Tr}}

\begin{document}

\title{Complex conjugation supermap of unitary quantum maps\\ and its universal implementation protocol}

\author{Jisho Miyazaki}
\affiliation{Department of Physics, Graduate School of Science, The University of Tokyo, Tokyo, Japan}

\author{Akihito Soeda}
\email{soeda@phys.s.u-tokyo.ac.jp}
\affiliation{Department of Physics, Graduate School of Science, The University of Tokyo, Tokyo, Japan}

\author{Mio Murao}
\affiliation{Department of Physics, Graduate School of Science,	The University of Tokyo, Tokyo, Japan}
\affiliation{Institute for Nano Quantum Information Electronics, The University of Tokyo, Tokyo, Japan}

\date{\today}

\begin{abstract}
A complex conjugation of unitary quantum map is a second-order map (supermap) that maps a unitary operator $U$ to its complex conjugate $U^*$.
First, we present a deterministic quantum protocol that universally implements the complex conjugation supermap when we are given a blackbox quantum circuit, guaranteed to implement some unitary operation, whose only known description is its dimension. 
We then discuss the complex conjugation supermap in the context of entanglement theory and derive a conjugation-based expression of the $G$-concurrence.  Finally, we present a physical process involving identical fermions from which the complex conjugation protocol is derived as a simulation of the process using qudits.
\end{abstract}

\maketitle

\paragraph{Introduction.---}
Limits of quantum information processing are drawn by the limits of quantum operations.  Every quantum operation is described by a mathematical map, but mathematically well-defined maps and implementable quantum operations are not equivalent.  This distinction stems from a fundamental fact that possessing a sample of quantum object is not the same as knowing its classical description, leading to various no-go theorems in quantum information~\cite{Park1973,DIEKS1982271,Wootters1982,KumarPati2000,PhysRevLett.98.080502}.  

While quantum operations on quantum states correspond to ``first-order'' maps defined on density matrices or vectors, maps can also be defined between these first-order maps and more generally for maps of any order~\cite{PhysRevA.72.062323,Gutoski:2007:TGT:1250790.1250873}.  These higher-order maps are collectively referred to as ``supermaps'' in Refs.\,\cite{Chiribella2008:architecture,Chiribella2008:supermaps,Chiribella2009}.
Completely-positive (CP) maps are a first-order map, which are realizable as a quantum gate within the standard quantum circuit model.
These gates may be provided as an input to a larger quantum protocol, which uses the input gates as a quantum subroutine.
The resulting operation implemented by the protocol depends on the input quantum gate.
Effectively, the protocol realizes a ``higher-order'' quantum operation, converting one quantum operation to another.

Universal implementations of supermaps assume little or no prior knowledge on the input quantum operation.
Generally, the supermaps whose universal implementation has an immediate application are often the ones impossible to implement universally, \textit{e.g.},
``cloning''~\cite{Chiribella2008:architecture,Chiribella2008:unitarycloning}, ``controllization''~\cite{Araujo2014:nogocontorol,Friis2014:controlunknowngate,PhysRevLett.114.190501,PhysRevA.94.022340,ChiribellaEbler2016,PhysRevA.95.062106,Thompson2018}, and ``quantum switch''~\cite{Colnaghi2012,Chiribella2013:quantumswitch,PhysRevLett.113.250402,PhysRevA.86.040301,PhysRevLett.120.120502}.  The inversion supermap $U \mapsto U^\dag$ is also known to have an application in quantum control~\cite{1602.07963,PhysRevX.8.031008}, but proven to be impossible~\cite{ChiribellaEbler2016}.
The no-go results for these supermaps hold under an additional assumption that the dimension of the input unitary operation is given.

\begin{figure}
	\centering
	\includegraphics[width=0.33\textwidth]{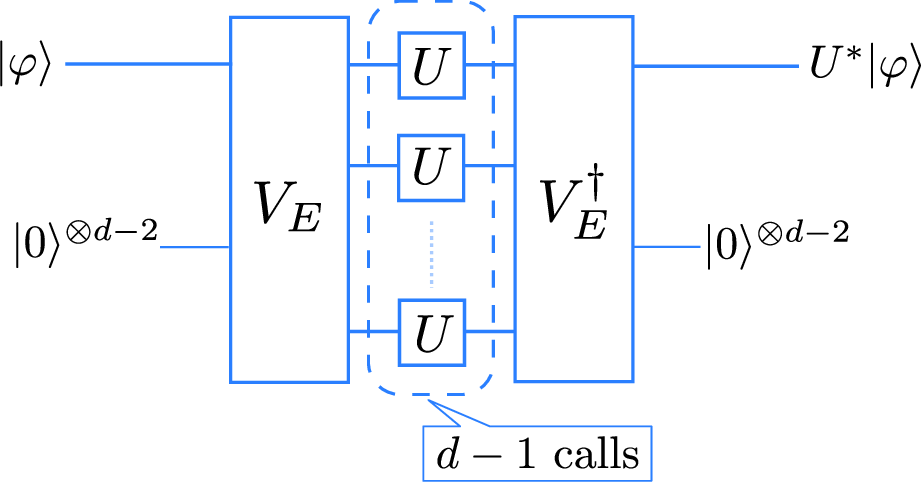}		\caption{\label{fig:algorithm}(color online)	The quantum circuit architecture to implement the universal unitary conjugation of a $d \times d$ unitary gate blackbox $U$, used $d-1$ times.  The circuit starts with $d-1$ qudits, each $d$-dimensional.  The first (top) qudit holds the input state and the rest are initialized in $\ket[0]$. Boxes labeled $V_E$ and $V_E^\dag$ are a quantum gate, defined in Eq.\,\eqref{V_E}.}
\end{figure}

The above inversion supermap inverts \textit{all} unitary operations.
An arguably less demanding supermap is a complex conjugation on unitary operators, defined by
\begin{equation} \label{unitary_cc}
 U = \sum u_{kl} \ket[k]\bra[l] \mapsto U^\ast = \sum u_{kl}^\ast \ket[k]\bra[l]
\end{equation}
with respect to some basis $\{\ket[k]\}$, which achieves an inversion for unitaries diagonal in the basis $\{\ket[k]\}$.
Although not quite the full unitary inversion, a deterministic universal complex conjugation of unitary operation leads to a probabilistic universal implementation of the full inversion whose failure probability decreases exponentially with the number of input unitary operations used~\cite{1810.06944}.

Unlike the full inversion, there exists a universal implementation of unitary complex conjugation that is also deterministic for $2 \times 2$ unitaries (see Eq.\,\eqref{su2conj}).
For larger dimensions, however, universal unitary complex conjugation is again unimplementable~\cite{ChiribellaEbler2016}, assuming that the implementation is deterministic and the input unitary operation is used only once.
This no-go result does not apply to \textit{approximate} implementations.  Reference~\cite{PhysRevA.81.032324} presents an approximate universal implementation of unitary complex conjugation that is optimal under a certain figure of merit.  Its approximation error improves with each additional use of the input operation, but an exact implementation requires an infinite number of uses.

A mathematical formulation of universally implementable of second-order maps is structurally similar to that of universally implementable first-order maps, \textit{i.e.}, the standard maps on quantum \textit{states}~\cite{Gutoski:2007:TGT:1250790.1250873,Chiribella2008:architecture,Chiribella2008:supermaps,Chiribella2009}.
As we discuss below, the complex conjugation supermap on unitary quantum maps is a complex conjugation on the corresponding Choi-Jamio{\l}kowski operators of the input maps.
In other words, a universal implementation of quantum \textit{state} complex conjugation immediately leads to that of unitary complex conjugation.  The former, however, is not admissible~\cite{Yang2014}.

In this Letter, we study the unitary complex conjugation supermap and its universal implementability as a higher-order quantum operation.
We first review the mathematical formulation of supermaps.
Despite the no-go~\cite{Yang2014}, we present a universal quantum algorithm that deterministically complex conjugates unitary quantum operations and argue how the no-go result is avoided.
The existence of universal implementability depends on the choice of target supermap but also on the set of input maps, thus universal implementablility of supermaps is an inherent property of the particular set of input maps.
For the unitary complex conjugation, we relate its universal implementability to entanglement theory, in particular, the entanglement measure of concurrence.  
Finally, we describe a physical process corresponding to the complex conjugation algorithm to offer a physical intuition behind the algorithm.

\paragraph{Universal complex conjugation of unitaries and states.---}
A first-order map, \textit{i.e.,} from states to states, is universally implementable if and only if the map is complete positive when expressed as a linear map on density matrices.
Implementability of supermaps is partially determined by implementability of maps on quantum states.
The Choi-Jamio{\l}kowski (CJ) isomorphism~\cite{Choi1975,JAMIOLKOWSKI1972275} establishes a duality between quantum operations and quantum states.
Let $\hil$ and $\hilk$ be two Hilbert spaces of dimension $d$, with bases $\ket[k]_\hil$ and $\ket[l]_{\hilk}$, respectively.
The Hilbert space $\hilk$ serves as a ``reference'' space of $\hil$.
Given a CP map $\Lambda$ from linear operators on Hilbert space $\hil$ to linear operators on Hilbert space $\hil'$, its CJ operator $\tilde{\Lambda}$ is an operator on $\hil' \otimes \hilk$ defined as
\begin{equation}
  \tilde{\Lambda} := \sum_{k,l=1}^d \Lambda(\ket[k]_\hil\bra[l]) \otimes \ket[k]_{\hilk} \bra[l].
\end{equation}
If a supermap on $\Lambda$ is implemented within the circuit model, then there exists a CP map on $\tilde{\Lambda}$.
Conversely, if a CP map corresponds to a given supermap, then it is implementable within the circuit model~\cite{PhysRevA.72.062323,Gutoski:2007:TGT:1250790.1250873,Chiribella2008:supermaps}, perhaps not deterministically, but heralded so that the successful instances are signaled. 
These facts imply that if the first-order map of universal state conjugation is CP, then a universal \textit{unitary} complex conjugation is implementable as a heralded and probabilistic quantum algorithm.
Nevertheless, the universal state conjugation violates CP.

Strictly speaking, the above CP condition on implementable supermaps assumes that the input quantum operation is used only once in the implementation circuit and that the supermap is defined for all CP maps including those not necessarily trace-preserving.
In general, multiple uses of the same input quantum operation may be possible, in which case the CP condition does not directly apply.
Moreover, CJ operators for unitary maps are rank 1, but CJ operators for general CP maps may have a higher rank.
However, universal state conjugation on pure states (hence, their density matrix is rank 1) is proven to be impossible, even under the relaxed condition of heralded probabilistic implementations with multiple but finite samples of the input quantum state~\cite{Yang2014}.

\paragraph{Implementation of universal unitary complex conjugation.---}
The input unitary is given as a quantum gate oracle implementing some unitary $U$.
The quantum circuit of our universal unitary conjugation algorithm is given in Fig.\,\ref{fig:algorithm}.
The algorithm starts with $d-1$ qudits, each of dimension $d$.
The qudits are labeled from $1$ to $d-1$ with the corresponding $d$-dimensional Hilbert spaces from $\hil_1$ to $\hil_{d-1}$, respectively.
The orthonormal basis vectors of each Hilbert space are $\ket[0],\ldots,\ket[d-1]$.  The choice of this basis decides the basis with which the unitary complex conjugation supermap is defined.  In what follows, we choose this particular basis for any $d$-dimensional Hilbert space.
The Hilbert space to which a given state belongs should be apparent from the context, but if necessary, we append a subscript as in $\ket[\varphi]_1 \in \hil_1$.
When completed, the algorithm applies $U^\ast$ on qudit $1$.
The remaining $d-2$ qudits are used as auxiliary systems, which are initialized to the state $\ket[0]$.
The first gate in the algorithm applies a unitary operator $V_E$ (defined below) on all the qudits.
Then, the input unitary operation $U$ is applied individually on each qudit.  This step requires $d-1$ calls of the input unitary operation in total.
Finally, $V_E^\dagger$ is applied, after which the state of qudit 1 results in the state with $U^\ast$ applied to its initial state, while the remaining qudits return to $\ket[0]$.

To define $V_E$, we introduce an isometric operator $E$ from $\hil_1$ to $\hil_1 \otimes \cdots \otimes \hil_{d-1}$, defined as
\begin{equation}
E := \sum_{\vec{k}} \frac{\epsilon_{\vec{k}}}{\sqrt{(d-1)!}} \ket[k_2,\ldots,k_{d}] \bra[k_1],
\label{eq:definition of antisymmetrizer}
\end{equation}
where $\vec{k} \in \{0,\ldots,d-1\}^{d-1} $ and $\epsilon_{\vec{k}}$ is the antisymmetric tensor of rank $d$.  We adopt the shorthand notation $\ket[k_2,\ldots,k_d] = \ket[k_2]_1 \otimes \cdots \otimes \ket[k_d]_{d-1}$.  The operator $E$ is an isometry, since for any $\ket[\varphi], \ket[\psi] \in \hil_1$, $\bra[\varphi]E^\dagger E \ket[\psi] = \braket{\varphi}{\psi}$.
Therefore, there exists a \textit{unitary} matrix $V_E$ on $\hil_1 \otimes \cdots \otimes \hil_{d-1}$ such that
\begin{equation} \label{V_E}
 V_E \ket[\varphi]_1\ket[\bar{0}]_{2 \cdots d-1} = E \ket[\varphi]_1,
\end{equation}
where $\ket[\bar{0}]_{2 \cdots d-1} := \ket[0]_2 \cdots \ket[0]_{d-1}$.

The correctness of the algorithm is guaranteed if 
\begin{multline} \label{correctness}
 V_E^\dag (U^{\otimes d-1}) V_E \ket[\varphi]_1 \ket[\bar{0}]_{2 \cdots d-1}
  = (U^\ast \ket[\varphi]_1) \ket[\bar{0}]_{2 \cdots d-1}.
\end{multline}
In terms of group representation theory, $E$ exploits the fact that the complex conjugate representation of $\SU[d]$ is unitarily equivalent to the antisymmetric subspace in the tensor representation of $\SU[d]$ on $\hil^{\otimes d-1}$.
First, let $\hilk[1]$, $\ldots$, $\hilk[d]$ be a $d$-dimensional Hilbert space.  We define an antisymmetric (unnormalized) state
\begin{equation} \label{antisym}
\ket[E] := \sum_{\vec{k}} \frac{\epsilon_{\vec{k}}}{\sqrt{(d-1)!}} \ket[k_1,\ldots,k_d] \in \hilk[1] \otimes \cdots \otimes \hilk[d].
\end{equation}
It is easy to see that $U^{\otimes d} \ket[E]$ is also an antisymmetric state for any $d \times d$ unitary.
Any antisymmetric state in $\hilk[1] \otimes \cdots \otimes \hilk[d]$ is proportional to $\ket[E]$.
For any $U \in \SU[d]$, this proportionality factor does not depend on $U$, in fact
\begin{equation} \label{inv}
U^{\otimes d} \ket[E] = \ket[E].
\end{equation}

To relate $\ket[E]$ to $E$, we interpret $\ket[E]$ as an operator from the $1$-dimensional Hilbert space $\mathbb{C}^1$ to $\hilk[1] \otimes \cdots \otimes \hilk[d]$.  In the following, we denote the identity operator on a given Hilbert space $\hil$ with a subscript as $I_{\hil}$.  Then, $I_{\hilb[1]} \otimes \ket[E]$ is an operator from $\hilb[1]$ to $\hilb[1] \otimes \hilk[1] \otimes \cdots \otimes \hilk[d]$.
Let $\bra[\Phi^{(d)}] :=  \sum_{k=1}^d \bra[k]\bra[k]$ be an operator from $\hilb[1] \otimes \hilk[1]$ to $\mathbb{C}^1$, which satisfies for any $U' \in SU(d)$
\begin{equation} \label{transpose}
 \bra[\Phi^{(d)}] (I_{\hilb[1]} \otimes U') = \bra[\Phi^{(d)}] ((U')^T \otimes I_{\hilk[1]}),
\end{equation} 
where $(U')^T$ is the transpose of $U'$ in the basis $\ket[k]$.
Next, define $J := \bra[\Phi^{(d)}] \otimes I_{\hilk[2] \cdots \hilk[d]}$.
The product $J  (I_{\hilb[1]} \otimes \ket[E])$ is an operator from $\hilb[1]$ to $\hilk[2] \otimes \cdots \otimes \hilk[d]$.  Thus, if we reinterpret $\hilk[j]$ as $\hilb[j-1]$, then $J  (I_{\hilb[1]} \otimes \ket[E])$ is equivalent to $E$.  
Therefore,
\begin{multline} \label{steps}
U^{\otimes d-1} E \simeq J (I_{\hilb[1]} \otimes (U^\dag U \otimes U^{\otimes d-1} \ket[E]))\\
= J (U^\ast \otimes (U \otimes U^{\otimes d-1} \ket[E]))  \simeq E U^\ast,
\end{multline}
where the first $\simeq$ follows from $U^{\otimes d-1} J = J (I_{\hilb[1]\hilk[1]} \otimes U^{\otimes d-1})$ and $U^\dag U = I$,
 the first equality from Eq.\,\eqref{transpose} with $U' = U^\dag$ and $(U^\dag)^T = U^\ast$, and the second $\simeq$ from Eq.\,\eqref{inv}.
Equation~\eqref{steps} and Def.\,\eqref{V_E} show that 
\begin{multline}
U^{\otimes d-1} V_E \ket[\varphi]_1 \ket[\bar{0}]_{2 \cdots d-1} =  U^{\otimes d-1} E \ket[\varphi]_1\\
= E U^\ast \ket[\varphi]_1 = V_E (U^\ast \ket[\varphi]_1) \ket[\bar{0}]_{2 \cdots d-1}.
\end{multline}
Finally, multiplying $V_E^\dagger$ from the left to both sides of this equation and using the unitarity of $V_E$ lead to Eq.\,\eqref{correctness}, which proves the correctness of the algorithm.

With respect to the no-go~\cite{Yang2014}, the only difference in our protocol is that the CJ operators for unitary maps satisfy an extra constraint, namely, the normalization $\tr_\hil \tilde{\Lambda} = I_{\hilk}$.
This constraint alone allows a universal implementation with an additional benefit of being deterministic.

\paragraph{Unitary conjugation and entanglement measure.---}
Entanglement is a property of quantum states, formally defined as correlations present in multi-partite quantum states which do not increase under the local operations and classical operations (LOCC)~\cite{PlenioVirmani2007,RevModPhys.81.865}.  The properties of LOCC determine which feature of quantum states qualifies as entanglement and how entanglement is affected by LOCC.

Local unitary operations are reversible, hence any function of quantum states, if it were to quantify entanglement, must be invariant under local unitary transformations. 
This is true for the \textit{concurrence} for two-qubit pure states $\ket[\psi] = \sum_{k,l=0}^1 c_{kl} \ket[k]\ket[l]$ defined in Refs.\,\cite{HillWootters1997,Wootters1998} as
\begin{equation}
\concurrence(\ket[\psi]) := |\bra[\psi] \sigma_y \otimes \sigma_y \ket[\psi^\ast] |
\end{equation}
with the Pauli $Y$ operator $\sigma_y = -i\ket[0]\bra[1]+i\ket[1]\bra[0]$ and $\ket[\psi^\ast]$ the complex conjugate of $\ket[\psi]$ in the basis $\ket[k]\ket[l]$.

The definition of $\concurrence$ is extended to mixed states via convex roofs.  This requires to find the set $\mathcal{S}_\rho$ of pure-state ensembles $\{p_k, \ket[\psi_k] \}_k$ that are consistent with the given mixed state, so that $\rho = \sum_k p_k \ket[\psi_k]\bra[\psi_k]$.
Then, $\concurrence(\rho)$ is defined as the minimum average pure-state concurrence over $\mathcal{S}_\rho$, \textit{i.e.,} $\concurrence(\rho) = \min_{\mathcal{S}_\rho} \sum_k p_k \concurrence(\ket[\psi]_k)$.
In general, the set of such ensembles possesses little mathematical structure to facilitate computing an accurate value of any measure defined via convex roofs.
The two-qubit concurrence is an exception in that the necessary optimization problem is already solved in Ref.\,\cite{Wootters1998}.
Analysis from Ref.\,\cite{Uhlmann2000} indicates that the use of complex conjugation appears to be a key mathematical property that allows the two-qubit concurrence to be solved for mixed states.

The concurrence $\concurrence$ has been generalized to higher-dimensional bipartite systems.  One such is the $G$-concurrence~\cite{Gour2005}
\begin{equation} \label{Gconc}
\concurrence^G(\ket[\psi]) = \alpha (\lambda_1 ... \lambda_d)^{1/d},
\end{equation}
where $\alpha$ is a normalization factor and $\lambda_k$ are the Schmidt coefficients of $\ket[\psi]$, \textit{i.e.}, 
$\ket[\psi] = (U' \otimes V' )\sum_k \sqrt{\lambda_k} \ket[k] \otimes \ket[k]$
for some local unitary $U'$ and $V'$.  The $G$-concurrence follows the analysis given in Ref.\,\cite{Rungta2001}, which generalizes $\concurrence$ to $\sqrt{\frac{d}{d-1}(1-\trace[\rho_\mathrm{r}^2])}$ from the reduced density matrix $\rho_\mathrm{r}$ of $\ket[\psi]$.

The local unitary invariance of $\concurrence$ is guaranteed from the fact that $\sigma_y$ achieves unitary complex conjugation for any $U \in \SU[2]$, \textit{i.e.},
\begin{equation} \label{su2conj}
 \sigma_y^{-1} U \sigma_y = U^\ast.
\end{equation}
To see this, observe that Eq.\,\eqref{su2conj} is equivalent to $\sigma_y = U^\dag \sigma_y U^\ast$, thus
\begin{multline}
 \concurrence((U \otimes V)\ket[\psi]) = |\bra[\psi] (U \otimes V)^\dag (\sigma_y \otimes \sigma_y) (U \otimes V)^\ast \ket[\psi^*]|\\
 = |\bra[\psi](\sigma_y \otimes \sigma_y) \ket[\psi^*]| =  \concurrence(\ket[\psi])
\end{multline}
for any $U, V \in \SU[2]$.  From our previous discussion, the unitary complex conjugation for higher-dimensional systems is possible by
\begin{equation}
 U^\ast = E^{-1} U^{\otimes d-1}E,
\end{equation}
where $E^{-1}$ is the generalized inverse (the Moore-Penrose inverse) of $E$.
Thus, we obtain a local unitary invariant generalization of $C$,
\begin{equation}
\concurrence^g(\ket[\psi]) := |\bra[\psi]^{\otimes d-1}(E \otimes E) \ket[\psi^\ast]|,
\end{equation}
which is a conjugation-based quantity much like the original concurrence.
The nonzero elements of the antisymmetric tensor $\epsilon_{\vec{k}}$ are for $\vec{k}$ such that all its elements differ, hence $\concurrence^g(\ket[\psi]) \propto \lambda_1 \cdots \lambda_d$.
Therefore, $\concurrence^G$ and $C^g$ are equivalent.

\begin{figure}[t]
	\centering
	\includegraphics[width=0.35\textwidth]{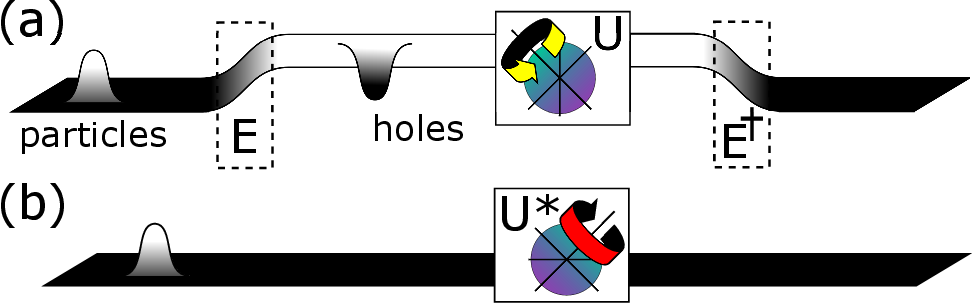}		\caption{\label{fig:particle_hole}(color online)	A conceptual figure of the particle-hole interpretation of the unitary conjugation algorithm.  The black and white strip represent the vacuum and fully occupied state of $d$-mode fermions.  The upward protrusion represents a single fermionic particle, while the downward indentation indicates a hole.  The processes $E$ and $E^\dagger$ in (a) correspond to a particle-hole exchange.  The hole in (a) undergoes the mode transformation $U$.  The lower figure (b) is an equivalent process to (a), in which the fermionic particle undergoes the mode transformation $U^\ast$.}
\end{figure}

\paragraph{Particle-hole interpretation.---}
For a supermap to lead to interesting applications, the action of the supermap must be nontrivial and admit an efficient universal implementation.  These two conditions are seldom satisfied because mathematically valid maps and physically implementable transformations do not coincide in quantum theory.
This gap between desired maps and implementable transformations is common in various areas of quantum information.
On the other hand, physical processes inherently realize a well-defined map that is guaranteed to be implementable.
Grover explains~\cite{doi:10.1119/1.1359518} that it was by analyzing a quantum diffusion process that lead to the discovery of his seminal search algorithm~\cite{Grover:1996:FQM:237814.237866}.
We shall see below that our universal complex conjugation algorithm follows quite naturally from a physical process in a fermionic system.

A system with $d$ fermionic modes is characterized by operators $a_k$ and $a_k^\dag$ for $k = 1,\ldots,d$, that obey anticommutation relations, $\{a_k, a_l\}=\{a_k^\dagger, a_l^\dagger\}=0$ and $\{a_k, a_l^\dagger\}=\delta_{kl}$.  The operator $a_k^\dag$ creates a fermionic particle of mode $k$, while $a_k$ annihilates it.  Denoting the vacuum state by $\ket[\vac]$, we define the completely occupied state $\ket[\full] := a_1^\dag a_2^\dag \cdots a_d^\dag \ket[\vac]$.  The action of $a_k$ on $\ket[\full]$ creates a ``hole'' in the completely occupied state, \textit{i.e.},
$a_k\ket[\full] = (-1)^{k-1} a_1^\dag \cdots a_{k-1}^\dag a_{k+1}^\dag \cdots a_d^\dag \ket[\vac]$.
We interpret this as a state with a single fermionic hole of mode $k$ whose corresponding creation operator is 
\begin{equation}
b_k^\dag := (-1)^{k-1} a_1^\dag \cdots a_{k-1}^\dag a_{k+1}^\dag \cdots a_d^\dag.
\end{equation}
Lastly, the effect of a unitary transformation $U$ on a fermionic particle is expressed by substituting the initial creation operator $a_k^\dag$ of each mode with $a_{U,k}^\dag := U a_k^\dag U^\dag$.

The physical process that simulates our unitary complex conjugation is as follows (see Fig.\,\ref{fig:particle_hole}).
First, a $d$-mode fermionic particle undergoes the particle-hole exchange $a_k^\dag \rightarrow b_k^\dag$.
Then, applied on the new hole is a fermion number preserving transformation $U = \exp(i H)$, where $H = \sum_{k,l=1}^d h_{kl} a_k^\dag a_l$ with $h_{kl}$ being the $(k,l)$-element of a Hermitian matrix $h$.  This achieves $b_k^\dag \rightarrow b_{U,k}^\dag = \sum_{l=1}^d u_{kl}^\ast b_l^\dag$, where $u_{kl}$ is the $(k,l)$-element of the unitary matrix $u = \exp(i h)$.  Note that $h$ may be any $d \times d$ Hermitian matrix, hence $u$ is an arbitrary $d \times d$ unitary matrix.  Finally, the resulting hole undergoes another particle-hole exchange $b_k^\dag \rightarrow a_k^\dag$, transforming $b_{U,k}^\dag$ to
\begin{equation}
{a'}_{U^\ast, k}^\dag := \sum_{l=1}^d [U]^*_{kl} a_l^\dag = U^* a_k^\dag (U^*)^\dag,
\end{equation}
where $U^\ast = \exp\left(-i\sum_{k,l=1}^d (h_{kl})^* a_k^\dag a_l\right)$.  All in all, the effect on the particle is mode transformation $U^*$.

Our fermionic hole is composed of $d-1$ fermionic particles.  Thus, a single $d$-level fermionic particle is simulated by a $d$-level qudit and a single fermionic hole by antisymmetric states of $d-1$ of such qudits.  The transformation $E$ in Eq.\,\eqref{eq:definition of antisymmetrizer} achieves precisely the particle-hole exchange in this qudit-simulation of the fermions.

\textit{Conclusion.---}
The mathematical similarities between a universal complex conjugation of quantum states and that of quantum operations may suggest that the known impossibility of the former forbids any implementation of the latter.
Nevertheless, we presented a deterministic quantum protocol that implements a universal complex conjugation of unitary operations as a higher-order quantum operation.
The action of unitary complex conjugation is analyzed in the context of entanglement theory, from which we derived an alternative expression of the $G$-concurrence using complex conjugation of states.
Finally, we described a physical process involving $d$-mode identical fermions that offers a physical interpretation of our complex conjugation protocol.

\begin{acknowledgments}
We thank Q. Dong, M.T. Quintino, A. Shimbo, and H. Yamasaki for helpful discussions.
This work was supported by JSPS KAKENHI Grant Number 15J11531, 26330006, 15H01677, 16H01050, 17H01694, and 18K13467.
\end{acknowledgments}

\end{document}